\documentclass[aps,prl,superscriptaddress,reprint,twocolumn,showpacs,amsmath,amssymb,floatfix,nofootinbib]{revtex4-1}
\usepackage{amsthm,graphicx,color}

\newcommand{\Om}{\Omega}
\newcommand{\omz}{\omega_z}

\newcommand{\rb}{{\bf r}}

\newcommand{\mT}{\mathcal T}
\newcommand{\mV}{\mathcal V}
\newcommand{\mW}{\mathcal W}
\newcommand{\Vt}{V_{\text{trap}} }

\begin{document}
\title{Onset and Irreversibility of Granulation of Bose-Einstein condensates under Feshbach Resonance Management}
\author{A. U. J. Lode$^{6}$}
\email{auj.lode@gmail.com}
\affiliation{Institute of Physics, Albert-Ludwig University of Freiburg, Hermann-Herder-Stra{\ss}e 3, D-79104 Freiburg, Germany}

\author{M. C. Tsatsos$^{6}$}
\email{mariostsatsos@gmail.com}
\affiliation{Honest AI Ltd., 65 Goswell Road, London EC1V 7EN, United Kingdom}
\affiliation{S\~ao Carlos Institute of Physics, University of S\~ao Paulo, PO Box 369, 13560-970, S\~ao Carlos, SP, Brazil}

\author{P. G. Kevrekidis}
\affiliation{Department of Mathematics and Statistics, University of Massachusetts, Amherst, Massachusetts 01003-4515 USA}

\author{G. D. Telles}
\affiliation{S\~ao Carlos Institute of Physics, University of S\~ao Paulo, PO Box 369, 13560-970, S\~ao Carlos, SP, Brazil}

\author{D. Luo}
\author{R. G. Hulet}
\affiliation{Department of Physics and Astronomy, Rice University, Houston, Texas 77005, USA}

\author{V. S. Bagnato}
\affiliation{S\~ao Carlos Institute of Physics, University of S\~ao Paulo, PO Box 369, 13560-970, S\~ao Carlos, SP, Brazil}

\date{\today}
\begin{abstract}
Granulation of quantum matter -- the formation of persistent small-scale patterns -- is realized in the images of quasi-one-dimensional Bose-Einstein condensates perturbed by a periodically modulated interaction. Our present analysis of a mean-field approximation suggests that granulation is caused by the gradual transformation of phase undulations into density undulations. This is achieved by a suitably large modulation frequency, while for low enough frequencies the system exhibits a quasi-adiabatic regime. We show that the persistence of granulation is a result of the irregular evolution of the phase of the wavefunction representing an irreversible process. Our model predictions agree with numerical solutions of the Schr\"odinger equation and experimental observations. The numerical computations reveal the emergent many-body correlations behind these phenomena via the multi-configurational time-dependent Hartree theory for bosons (MCTDHB).
\end{abstract}

\maketitle

\footnotetext[6]{These authors contributed equally to the manuscript.}

{\it Introduction.} 
Conventional or classical granular matter is multifaceted and exhibits features of different phases~\cite{Jaeger1996,Mehta1994}. For instance, a granular system can behave as either gas, liquid, or solid depending on its configuration and interaction with the environment. 
Analogously, \textit{quantum} granular matter combines different quantum properties like quantum turbulence~\cite{henn:09,thompson:11,shiozaki:11,seman:11,navon:16,tsatsos:16,madeira:20} or the formation of fluctuating localized structures~\cite{GRAINS}. Ref.~\cite{GRAINS} adds granulation to the list of many-body properties of Bose-Einstein condensates (BECs) in dilute atomic gases~\cite{Pethick,Bogoliubov,PitaSandro}. This list has grown since the first production of BEC~\cite{BEC:Sodium,BEC:Lithium,BEC:Rubidium} and contains, for instance, also the Hubbard~\cite{CB98,BH} and Dicke~\cite{DickeOrg,CavityExp1,hamner:14} models, supersolids~\cite{supsol,supsol1,supsol2} and topological states~\cite{zhang:18,quantumHall,topo}.

Quantum granulation has been hinted at in experimental~\cite{seman:11} and theoretical results~\cite{Yukalov2014} and is a manifestation of quantum fluctuations in many-body systems far from equilibrium~\cite{GRAINS}. In Ref.~\cite{GRAINS}, granular states were created in a BEC by periodically modulating the interaction strength of the atoms~\cite{Pollack2010} with a sufficiently large amplitude and frequency for a sufficiently long duration. Here and throughout this work we consider the quantum case of granulation.

Remarkably, granulation entails the emergence of many-body correlations: it appears in single-shot images~\cite{Kaspar16,Axel17} but is not captured fully with the density, i.e., when many single-shot images are averaged~\cite{GRAINS}. In quantum granular matter the domains, where coherence is maintained and quantum fluctuations are small, are limited to small `islands', i.e. the grains of the system. 
During the formation of the granular state in an atomic BEC, the mean-field description~\cite{gross:61,pitaevskii:61} thus breaks down. This is rather natural to expect given the strong periodic driving.

For granulated states, the one-body reduced density matrix (1-RDM) $\rho^{(1)}(z,z')=\langle \Psi \vert \hat{\Psi}^\dagger(z) \hat{\Psi}(z') \vert \Psi \rangle=$ $\sum_k \alpha_k^{(1)} \phi_k^*(z',t)\phi_k(z,t)$ acquires
several macroscopic eigenvalues $\alpha_k^{(1)}$ and contributing eigenfunctions $\phi_k(z,t)$~\cite{GRAINS}. BECs that are described by a 1-RDM with multiple macroscopic eigenvalues are known as fragmented BECs~\cite{Spekkens99,Noizieres83,Penrose56,Mueller06}.
The breakdown of the mean-field Gross-Pitaevskii (GP) description of the condensate has been anticipated for granular states, because of the irregular phase profiles found in Ref.~\cite{Yukalov2014}. 
For a single-particle model, a periodic driving of the trap with a certain amplitude and frequency may transfer atoms from the ground state to a variety of excited states~\cite{Yukalov1997}.
Ref.~\cite{GRAINS} suggests that for modulated interactions, this finding is true for each of the eigenfunctions of the 1-RDM and that granulation is connected with the buildup of correlations. However, Ref.~\cite{GRAINS} does not explain the onset and irreversibility of granulation.

\begin{figure}[ht!]
	\includegraphics[width=1.0\columnwidth]{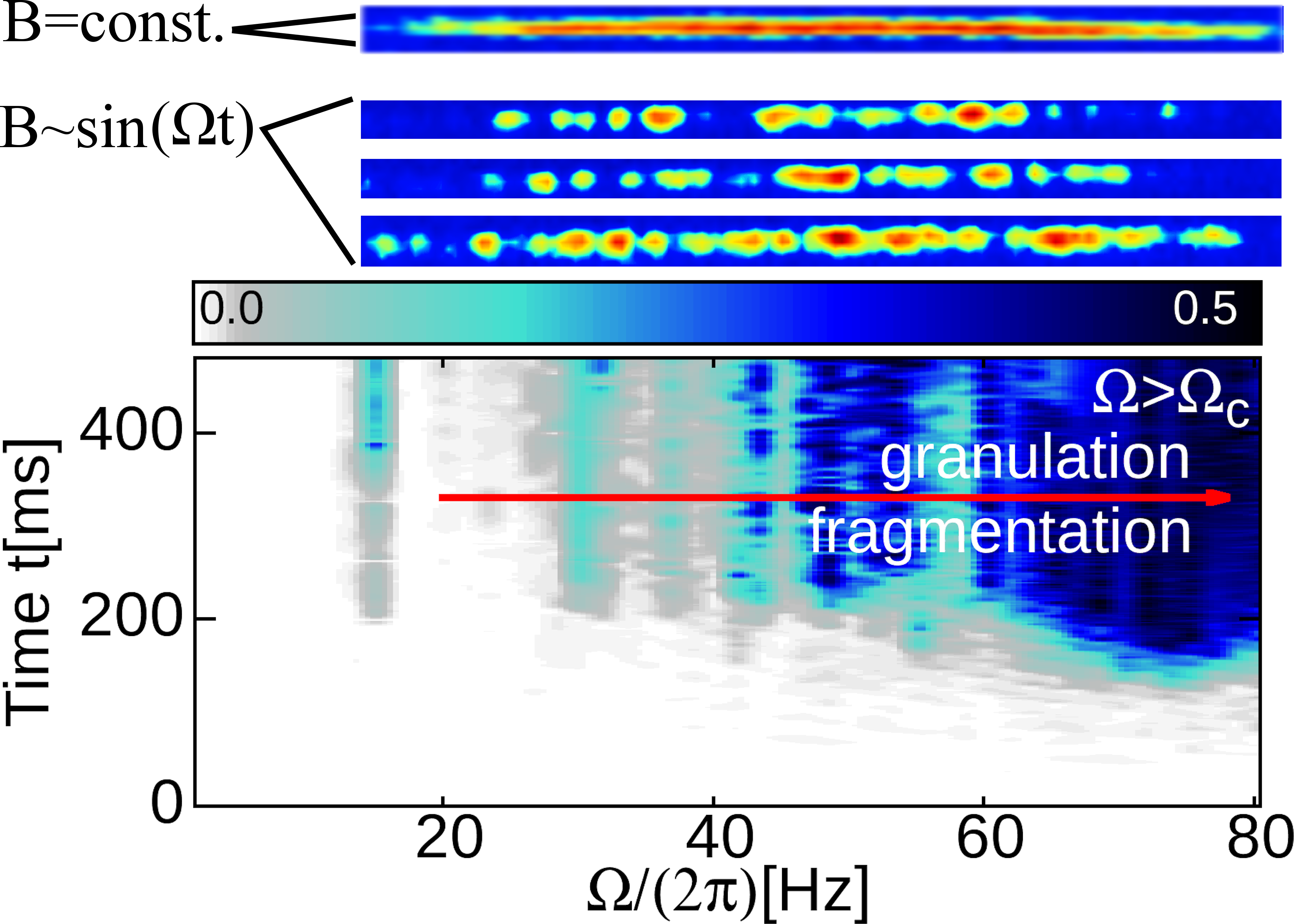}
	\caption{Time-periodic modulations of the magnetic field lead to the formation of a granulated state in a quasi-one-dimensional Bose-Einstein condensate. Top: experimental images of the initial and granulated many-body state (adapted from Fig.~6 of Ref.~\cite{GRAINS}). Bottom: the magnitude of the excited fraction $1-\sum_{k=2} \frac{\alpha_k^{(1)}}{N}$, i.e. the number of particles not occupying the first eigenfunction of the 1-RDM. Here, white is zero and black is half the particles. The excited fraction grows as granulation emerges for $\Omega> \Omega_c \approx (2\pi)30$Hz and the 1-RDM of the system attains multiple significant eigenvalues; granulation and fragmentation emerge together in agreement with the observations in Ref.~\cite{GRAINS}.}
	\label{Fig:SKETCH}
\end{figure}

In this Letter, we fill in this gap and investigate the mechanism behind the formation and irreversibility of granulation. We consider a BEC subject to a periodically modulated interparticle interaction strength (Fig.~\ref{Fig:SKETCH}, top).
We show that the granulation of quantum matter is accompanied by an
irregular, disordered behavior of the phase of the many-body
state which bears some features of quantum
turbulence~\cite{henn:09,thompson:11,shiozaki:11,seman:11,navon:16,tsatsos:16,madeira:20}. This irregularity, in turn,
renders the formation of grains in the system an irreversible process.
The irregular phase profile of the granulated many-body
state can be understood as the result of two interrelated processes.
\textit{One}, a naturally emergent modulation of the phase of \textit{each} of the single-particle wavefunctions (or \emph{orbitals}), that build up the fragmented many-body state; we analytically demonstrate this phase modulation using a Dyson series of the time-evolution operator. The phase of an initially condensed state rapidly oscillates from the start of the external perturbation and, at later times, the phase-modulation triggers the formation of dips in the density. \textit{Two}, an exponentially large number of single-particle phases contributes to the many-body phase as coherence is lost. The first of the above-mentioned processes takes place on the single-particle level, while the second one is a genuine many-body process.

We corroborate our analytical arguments about granulation by numerical simulations using the multiconfigurational time-dependent Hartree method for bosons (MCTDHB)~\cite{GRAINS,ultracold,MCTDHX,MCTDHB,MCTDHB2,MCTDHX1,MCTDHX2,lin:20,RMP:20}. We demonstrate that our analytical approach adequately captures the early stage of the formation of the grains via the emergence of jumps in the phase profiles. The density and phases of different orbitals as well as the correlation functions we find in our numerical computations underpin our analytical finding that the persistence of granulation is caused by an irregular evolution of the phase combined with the emergence of fragmentation.


{\it Setup.} We consider $N$ interacting bosons described by the time-dependent wavefunction $\Psi(z_1,...,z_N,t)$ whose dynamics is governed by the time-dependent Schr\"odinger equation $\hat{H}\vert \Psi \rangle = i \partial_t \vert \Psi \rangle$ with the Hamiltonian
\begin{equation}
H = \mT + \mV + \mW
\label{hamiltonian}
\end{equation}
and $\mT=-\frac{1}{2} \sum_i^N \nabla_{\rb_i}^2$,  $\mV=\sum_i^N \Vt(\rb_i)$ and $\mW=\sum_{i\neq j}W(\rb_i-\rb_j,t)$ being the $N$-particle kinetic, potential and interaction energy operators, respectively. We work with dimensionless units~\cite{units} and consider a gas confined by an elongated cigar-shaped trap with longitudinal (loose) axis $z$ -- our spatial coordinate -- and transverse (tight) axes $x,y$.
To achieve granulation of the state $\vert \Psi \rangle$, we use a harmonic external confinement, $\Vt = \frac{\omz^2}{2}z^2, \omz=0.1$ and contact interactions with a time-dependent interaction strength,
\begin{eqnarray}
W(z_i-z_j,t) &=& g(t) \delta(|z_i - z_j|);\label{TDint} \\
g(t)&=& g_0 \left[\beta_1  (-1 + (\beta_2 - \beta_3 \sin(\Om{}t))^{-1}\right].\nonumber
\end{eqnarray}
Such modulated interactions are the effect of a periodic oscillation of the magnetic field close to the Feshbach resonance, hence, the term Feshbach resonance
management~\cite{FRM}, as used in the experiment reported in Ref.~\cite{GRAINS}. Here we consider variations of $g(t)$ with an amplitude $\sim 0.9 g_0$, see Supplemental Material (SM)~\cite{SM} Sec.~S1 for the dependence of the parameters $\beta_j$ on the magnetic field and scattering properties of lithium.
Feshbach resonance
management has attracted a substantial amount of
attention, see~\cite{malomed,kartashov}. At the mean-field level, these works have highlighted interesting features of the process like the formation of breathing dynamics~\cite{FRM} and the potential stabilization of higher-dimensional bright solitons against collapse in attractive
condensates~\cite{saito:03}. However, when granulation of the state is observed as in~\cite{GRAINS}, the many-body character of the process cannot be overlooked.

We now specifically analyze the dynamics of $N=10,000$ bosons governed by the Hamiltonian in Eq.~\eqref{hamiltonian} for two cases: an interaction strength modulated at a frequency $\Om=2.5\omz$, below the threshold for the emergence of granulation, and an interaction strength modulated at a frequency $\Om=10\omz$, above said threshold. Here, and in the following we use $\omz=(2\pi) 8$~Hz. The modulation lasts for $t_m=250$~ms and we monitor the evolution for an additional hold time $t_h=250$~ms with our numerical solutions of the time-dependent many-boson Schr\"odinger equation.

\begin{figure*}[t]
	\mbox{\includegraphics[width=\linewidth]{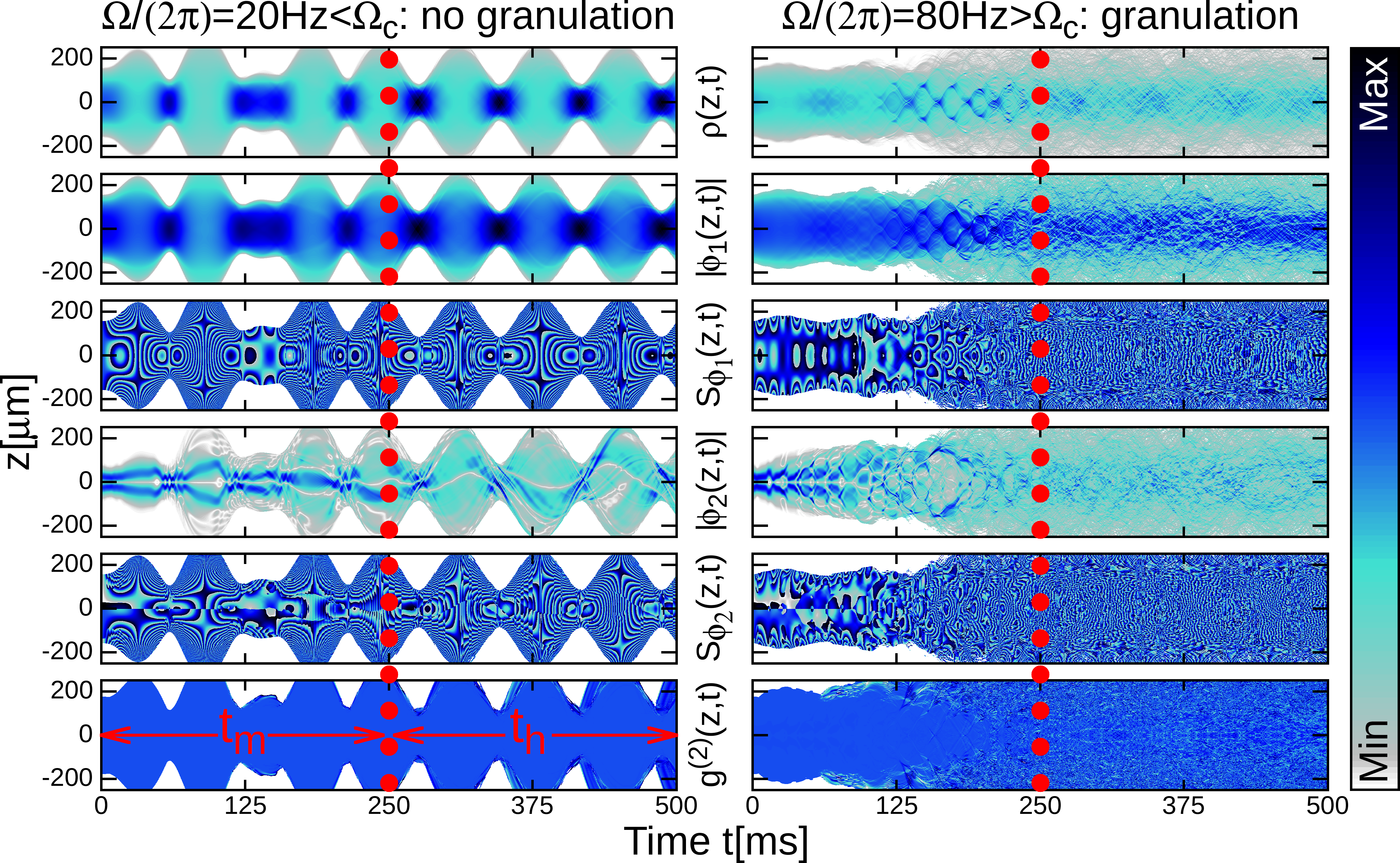}}
	\caption{Tracing adiabatic evolution and granulation in the state of a BEC.
		Left column: a quasi-adiabatic time-evolution of the density $\rho(z,t)$, the orbital densities $\vert \Phi_{1/2}(z,t)\vert$, their phases, $S_{\Phi_{1}}(z,t),S_{\phi_{2}}(z,t)$, and the coherence $\vert g^{(2)}(z) \vert^2$ are seen for a modulation of the interactions at a frequency $\Om=2.5\omz<\Omega_c$. Right column: for a modulation of the interactions at a frequency $\Om=10\omz>\Omega_c$ the state becomes granular and the same quantities as in the left column start to displaying irregular features. The vertical dotted red line at $t_m=250$ms shows where the modulation of the interactions is stopped and the holding time $t_h=250$ms starts (labels and arrows in bottom left panel).		
		The ranges of the color scale are different for the different quantities: $\rho(z,t)\in \left[0,\sim 0.1\right]; \vert \phi_{1}(z,t) \vert \in \left[0,\sim 0.1 \right],\vert \phi_{2}(z,t) \vert \in \left[0,\sim 0.1 \right], S_{\Phi_{1}}(z,t)\in\left[0,2\pi\right],S_{\phi_{2}}(z,t)\in\left[0,2\pi\right], \vert g^{(2)}(z,t)\vert \in \left[0.3,1.7\right]$. We excluded points $z$ from the plots of $\Phi_{1}\Phi_{2},S_{\Phi_{1}},S_{\Phi_{2}},$ and $g^{(2)}$ for which $\rho(z,t)<0.001$. Results on the density $\rho(z,t)$ are computed with the same parameters as Fig.~10 in Ref.~\cite{GRAINS}. See text for further discussion.}
	\label{Fig:GRAINS}
\end{figure*}

{\it Results.} At $\Om=2.5\omz<\Om_c$ granulation is absent and the time-evolution of the density $\rho(z,t)=\langle \Psi \vert \hat{\Psi}^\dagger(z) \hat{\Psi}(z) \vert \Psi \rangle$ is quasi-adiabatic: the gas follows the perturbation \emph{in phase} and the cloud retains its Thomas-Fermi profile with a time-oscillating radius and no structural changes (Fig.~\ref{Fig:GRAINS}, top left).
This quasi-adiabatic evolution is also seen in the orbital densities $\vert \phi_1(z,t) \vert$ and $\vert \phi_2(z,t)\vert$. The phases of the orbitals only display very slow spatio-temporal modulations and second-order coherence $g^{(2)}(z) = \frac{\langle \Psi \vert \hat{\Psi}^\dagger(z) \hat{\Psi}^\dagger(-z) \hat{\Psi}(z)\hat{\Psi}(-z) \vert \Psi \rangle}{\rho(z)\rho(-z)}$ is maintained ($\vert g^{(2)}(z)\vert^2 \approx 1$) over large distances (Fig.~\ref{Fig:GRAINS}, bottom left).

At $\Om=10\omz>\Om_c$ the BEC granulates. The time-evolution of the density $\rho(z,t)$ features breathing-like dynamics in this case, also. Additionally, short-scale undulations that emerge in regions of intermediate density values are seen. As time proceeds, these undulations gradually expand towards the center of the cloud (cf. Fig.~\ref{Fig:GRAINS}, right column). 
The behavior of the orbitals $\phi_1(z,t)$ and $\phi_2(z,t)$ (Fig.~\ref{Fig:GRAINS}, right) that build up the granular state follows that of the density $\rho(z,t)$. The state displays coherence $g^{(2)}\sim 1$ on a length-scale that shrinks with time. When the modulation has been switched off at times $t>t_m=250$ms the domains of coherence where  $g^{(2)}\sim 1$ are limited to a few small ``islands'', i.e. the grains of the system.

The irregular, noisy patterns in the quantities in Fig.~\ref{Fig:GRAINS}, right column, are a hallmark of granulation that causes the single-shot images of the system to have isolated clusters of particles with fluctuating size (cf. Ref.~\cite{GRAINS} and Fig.~\ref{Fig:SKETCH}, top). Even after the modulation of the interactions is turned off ($t>t_m=250$ms) the emergent patterns persist without decaying, for at least another $t_h=250$ms. We now turn to the mechanism behind the onset and persistence of granulation. Our analytical considerations show the crucial role played by the quasi-irregular patterns in the phases $S_{\phi_{1}},S_{\phi_{2}}$ and correlations $g^{(2)}$ of the granulated state in Fig.~\ref{Fig:GRAINS}. 

For the adiabatic time-evolution of an initial Thomas-Fermi density $\rho_{TF}(z,t)$, we approximate the modulated interparticle interaction [Eq.~\eqref{TDint}] with $g(t)\equiv g_0 (1+A\sin(\Omega t))$ and insert it in $\rho_{TF}(z,t)$:
\begin{eqnarray}
 \rho_{TF}(z,t) =& \frac{\mu - V(\vec{r})}{g} = \frac{\mu-\frac{1}{2}\omz^2 z^2}{g_0(1+A \sin(\Omega t))} \nonumber \\
 \simeq& \rho_{TF}(z,0) \left( 1- A \sin (\Omega t) \right), \label{order0}
\end{eqnarray}
the latter holding for small $A$. This adiabatic approximation yields a density oscillation of period $T\approx 50$~ms, which is in qualitative agreement with the observation that the observables in the left panel of Fig.~\ref{Fig:GRAINS} reach a maximal extent four -- five times for $0<t<250$~ms. For $t>250$~ms and no external drive the system oscillates with its natural breathing frequency and observables reach their maximal extent roughly every $50$~ms. Notably, in the case of granulation (Fig.~\ref{Fig:GRAINS} right) the orbital phases $S_{\phi_{1}}(z,t),S_{\phi_{2}}(z,t)$ for $t\lesssim125$ms oscillate at roughly half the modulation frequency  ($\sim 6$ cycles over $t=125$ms or $40$Hz in dimensional units). A response at half the driving frequency is a shared feature with the Faraday waves theoretically predicted in~\cite{Staliunas2004} and experimentally reported in~\cite{Engels2007,GRAINS,smits:18}. We note, that $M>2$ orbitals would be necessary for a quantitatively more accurate description, but are beyond current computational capabilities for $N=10000$ particles. We checked the reliability of our results for different $M$ and $N$ in Ref.~\cite{GRAINS} (see SM~\cite{SM} Sec.~S2 for other numerical details).

To approximate the non-adiabatic time-evolution with the Hamiltonian in Eqs.~\eqref{hamiltonian}--\eqref{TDint} we use a Dyson series. 
To leading order, the state at some time $t_1>t_0=0$ is: 
\begin{equation}
\phi(z,t_1) \approx \exp [-i \vartheta(z,0,t_1) ] \phi(z,0). \label{order1}
\end{equation}
Here, $\vartheta(z,t_0,t_1)$ is a function of the modulation frequency $\Omega$ and the density $\vert \phi(z,t_0) \vert^2$, see SM~\cite{SM} Sec.~S2 for details.
The modulation of the interactions thus leads to a modulation of the phase of the state, $\exp [-i \vartheta(z) (0,t_1) ]$ as it evolves from time $t=0$ to time $t_1$. 

The density at time $t_1$ is identical to the density at time $t_0$, because the phase factor $\exp [-i \vartheta(z) (0,t_1) ]$ in Eq.~\eqref{order1} cancels out, $\rho(z,t_1)=\vert \rho(z,0) \vert^2$.
To second order in the Dyson series at a time $t_2=t_1+\Delta t$ we find the density:
\begin{equation}
 \rho(z,t_2) \approx  \rho(z,0)\left[1+ T(z,\Delta t) \right].
 \label{order2}
\end{equation}
Here, the term $T(z)$ depends on the modulation frequency $\Om$, the density of the initial state, $\vert \rho (z,0) \rangle$, and its derivatives. Importantly, Eq.~\eqref{order2} is not just a phase factor, but determines the density modulation at time $t=t_2=t_1+\Delta t$; see SM~\cite{SM} Sec.~S2 for details about $T(z)\propto \frac{\partial }{\partial z}\vert  \phi(z,t_0) \vert^2 $. Eq.~\eqref{order2} highlights that the continued modulation of the interactions causes the phase modulations in the state at $t_1$ [Eq.~\eqref{order1}] to metamorphose into modulations of the density at times $t_2>t_1$. As the term $T(z)$ depends on the partial derivative of the density of the state, it makes the ensuing modulation of the density most pronounced where the gradient of the density is large. This observation is in agreement with our numerical results on the formation of the granular state in Fig.~\ref{Fig:GRAINS}. 

Our analytical model explains \textit{on a single-particle level} the quasi-adiabatic regime where the modulation of interactions causes a breathing-like oscillation without a density modulation [Eq.~\eqref{order0}]. Our model also describes how density undulations result from phase undulations [Eqs.~\eqref{order1} and \eqref{order2}] in the non-adiabatic, granulated regime. Granulation, however, is seen in single-shot images, but is not fully captured in their average, the density~\cite{GRAINS}. The irregular behavior seen in the phases and densities in the right panel of Fig.~\ref{Fig:GRAINS} shows an outstanding stability, not seen in other similar setups. For instance, the experiment~\cite{navon:16} has shown re-equilibration of a turbulent state of a BEC. Granulation, in contrast (Figs.~\ref{Fig:SKETCH},~\ref{Fig:GRAINS} and Ref.~\cite{GRAINS}) is \textit{non-transient} and survives for long times, without reversing to a non-granular state. This irreversible behavior is underpinned by the observation that the system has to be driven to a state very far from equilibrium in order to feature granulation; in three dimensions, an excitation beyond the quantum turbulent regime with a chaotic tangle of many vortices is required to reach the granular state~\cite{shiozaki:11,seman:11}. 

From a mean-field perspective, arguably, the closest analogue to the
granulated state is a coherent structure involving multiple density dips similar to dark solitons reviewed in~\cite{djf}.
However, a multi-dark-soliton ansatz approximating a GP granular state is prone to dynamical instabilities (e.g. Ref.~\cite{Coles2010}). We continue by highlighting evidence that many-body correlations are what makes the granular state robust dynamically: the depletion of a state en route to granulation grows in
time, ultimately resulting in a state with a large number of
grains~\cite{GRAINS}.
In that light, a mean-field wavefunction of the form $\vert \Psi \rangle = \vert N,t \rangle$ that assumes total condensation in one orbital cannot fully describe the system. 

We now consider analytically the ansatz we used in our numerical solutions of the Schr{\"o}dinger equation where the particles have the freedom to occupy either of $M=2$ \emph{time-dependent} orbitals,
\begin{equation}
\vert \Psi \rangle= \sum_{k=0}^N C_k (t) \vert N-k, k ,t \rangle. \label{M2}
\end{equation}
For such a state, the emergence of correlations means that multiple configurations $\vert N-k, k,t \rangle$ are contributing, i.e. several expansion coefficients $C_k(t)$ become significant with time. 
Our MCTDHB results show that correlations emerge (see Fig.~\ref{Fig:GRAINS}) and that correlations and granulation persist after the modulation has stopped.
We thus conclude that the formation of grains is an irreversible process for our driven many-body system.

We now discuss the connection of the phase and the symmetry of the many-body state [Eq.~\eqref{M2}] and draw conclusions for the irreversibility of the granulation process. 
The many-body phase $\Theta$ is connected to the phases of the single-particle states through Eq.~\eqref{M2} in a non-trivial way.
It is, however, straightforward to see (SM~\cite{SM} Sec.~S3) that every contributing configuration $\vert N-k, k, t\rangle$ in $\vert \Psi \rangle$ [Eq.~\eqref{M2}] contains exponentially many distinct products of single-particle phases. This is due to the nature of the complete bosonic symmetrization of the state $\Psi$: every basis state $\vert N-k,k,t \rangle$ is a sum of exponentially many products of orbitals.
The modulation of the interaction triggers phase modulations of the single-particle states [Eqs.~\eqref{order1} and \eqref{order2}], beginning directly with the start of the modulation of the interactions. As granulation emerges, fragmentation grows (Fig.~\ref{Fig:SKETCH}) and numerous coefficients $C_k$ become significant. With each significant coefficient, an exponentially large number of distinctly modulated phase terms is added to the many-body phase.

We infer that granulation is an irreversible process, because of this explosion of the number of contributing phase terms in the time evolution of the state: when a threshold complexity in the many-body phase $\Theta$ has been crossed, it cannot equilibrate any more. This threshold separates the two physically distinct cases of the quasi-adiabatic evolution (Fig.~\ref{Fig:GRAINS} left)
and the non-adiabatic evolution for the granular state (Fig.~\ref{Fig:GRAINS} right). Our results underpin, in accordance with Ref.~\cite{GRAINS}, that the granulation of BECs is a beyond-mean-field effect that cannot be fully covered within a GP description.
The irreversibility is also reflected in the entropies of the one- and two-body density matrices $\mathbf{S}_{\rho^{(1)}}$ and $\mathbf{S}_{\rho^{(2)}}$ \cite{chakrabarti:15,roy:18} of the state after the modulation. These entropies remain minimal in the adiabatic case and increase considerably only in the case of granulation (see Sec.~S4 in the SM~\cite{SM}).

The correlation function is directly related to the many-body phase $\Theta$~\cite{Stimming2011,Glauber}; see plots in the bottom panels of Fig.~\ref{Fig:GRAINS}. Indeed, for the quasi-adiabatic evolution at low modulation frequency, the created phase undulations of the single-particle states do not affect the dynamics of the coefficients of the state and its correlation function remains constant almost across the complete BEC: only a single contributing configuration $\vert N,0,t \rangle$ determines the evolution of the state. The latter effectively stays within the previously explored mean-field regime~\cite{FRM,malomed}. In contrast, for the granular case, several contributing configurations $\vert N-k, k ,t \rangle$ cause a complicated many-body phase and an irregular evolution of the correlation function (see SM~\cite{SM} Sec.~S5 for a verification for higher-order correlations). The irreversibility of the granulation process is manifest in the persistence of this irregular evolution of the correlations after the modulation is turned off at time $t_m=250$ms, see bottom right panel of Fig.~\ref{Fig:GRAINS}. 

{\it Conclusions.} We have shown that
the modulation of the interactions in Bose-Einstein condensates can provide a
systematic platform for exploring the phenomenon of quantum granulation
and its many-body character.
On the single-particle level, we used a perturbative expansion of the time-evolution operator to explain how the modulation of the interaction strength transforms first into phase and later into density undulations. 
The granular state
has been speculated to lose its phase coherence from the mean-field perspective in Ref.~\cite{Yukalov2014}; here we have highlighted that correlations are indispensable for its persistence and full many-body characterization. On the many-body level, phase undulations of single-particle states cause an exponentially growing complexity of the many-body phase and are responsible for the irreversibility of the granulation process: the correlations imprinted on the system persist even after the modulation of the strength of the interactions has been stopped.

Our work is a starting point for systematically exploring
granulated states, but numerous questions remain open: the long-time asymptotics of grains, their characteristic length scales (if applicable) and what controls them, the threshold for the emergence of granular states and its possible connection to Mathieu equations~\cite{pelster:14}, as well as the nature of the transition from non-granular to granular asymptotic configurations. 
Moreover, the connection of granulation to quantum turbulence~\cite{henn:09,thompson:11,shiozaki:11,seman:11,navon:16,tsatsos:16,madeira:20} and its interplay with the dimensionality of the system are of interest.

\begin{acknowledgments}
AUJL acknowledges financial support by the Austrian Science Foundation (FWF) under grant P32033-N32 and the Wiener Wissenschafts- und TechnologieFonds (WWTF) project No MA16-066. MCT acknowledges financial support from FAPESP. Computation time on the Hazel Hen cluster of the HLRS in Stuttgart is gratefully acknowledged.  This material is based upon work supported by the US
 P.G.K. acknowledges support of the National Science Foundation under Grants PHY-1602994	and DMS-1809074, as well as support of the Alexander von Humboldt Foundation. RGH acknowledges support from the NSF (Grant PHY-2011829) and the Welch Foundation (Grant C-1133).
 \end{acknowledgments}


\begin{thebibliography}{99}
\bibitem{units} In the experiment $\omega_r/\omz\approx32$ and so we can consider the gas being in the 1D limit. In the numerical simulations we divide the Hamiltonian for de-dimensionalization by $\frac{\hbar}{mL^2}$ ($m$ is the mass of lithium-$7$). To arrive at $\omz/(2\pi)=8$Hz and $\omega_r=2\pi 254$Hz we choose $L=4.3\mu$m. To emulate the scattering length close to $7.5 a_0$ for $3\cdot 10^5$ lithium atoms, we choose $g_0 (N-1)=357$. The values are thus chosen corresponding to the ones used in the experiment \cite{GRAINS} and imply a unit of time of $\frac{mL^2}{\hbar}=2$ms. See Sec.~S1 in the SM~\cite{SM} for further details on the presented numerical simulations.
\end{thebibliography}
\end{document}


\title{Supplemental material:\\Onset and Irreversibility of Granulation of Bose-Einstein condensates under Feshbach Resonance Management}
\author{A. U. J. Lode}
\email{auj.lode@gmail.com}
\affiliation{Institute of Physics, Albert-Ludwig University of Freiburg, Hermann-Herder-Stra{\ss}e 3, D-79104 Freiburg, Germany}

\author{M. C. Tsatsos}
\affiliation{Honest AI Ltd., 65 Goswell Road, London EC1V 7EN, United Kingdom}
\affiliation{S\~ao Carlos Institute of Physics, University of S\~ao Paulo, PO Box 369, 13560-970, S\~ao Carlos, SP, Brazil}

\author{P. G. Kevrekidis}
\affiliation{Department of Mathematics and Statistics, University of Massachusetts, Amherst, Massachusetts 01003-4515 USA}

\author{G. D. Telles}
\affiliation{S\~ao Carlos Institute of Physics, University of S\~ao Paulo, PO Box 369, 13560-970, S\~ao Carlos, SP, Brazil}

\author{D. Luo}
\author{R. G. Hulet}
\affiliation{Department of Physics and Astronomy, Rice University, Houston, Texas 77005, USA}

\author{V. S. Bagnato}
\affiliation{S\~ao Carlos Institute of Physics, University of S\~ao Paulo, PO Box 369, 13560-970, S\~ao Carlos, SP, Brazil}

\begin{abstract}
	In this Supplemental Material, we provide the details of our numerical simulations using the multiconfigurational time-dependent Hartree method for bosons, see Sec.~\ref{Sec:Numerics}.
	We analyze the time-evolution of a system of ultracold atoms subject to a time-periodic modulation of interactions using a Dyson series for the time-evolution operator in Sec.~\ref{Sec:Dyson1}. We find that the time-dependent interactions first result in a modulation of the phase that is later transmogrified into a modulation of the density. 
In Sec.~\ref{Sec:Phase}, we show that the phase of a many-body state accumulates contributions from exponentially many distinct one-body phases for every configuration contributing to its wavefunction and provide complementary numerical results on the entropy and high-order correlations embedded in the many-body state in Sec.~\ref{Sec:Entropy} and Sec.~\ref{Sec:HC}, respectively. 
\end{abstract}

\date{\today}

\maketitle

\section{Simulation details} \label{Sec:Numerics}

\subsection{Interaction model}
The values of the parameters in our simulations are selected in accordance with magnetic field,
\begin{equation}
B(t)=\bar{B}+\Delta \sin(\Omega t), \label{eq:B}
\end{equation}
used in the experiment in Ref.~\cite{GRAINS} with ultracold lithium atoms.
Due to the Feshbach resonance of lithium-$7$ at $B_{\infty}$, the 
scattering length of the atoms changes with time as follows:
\begin{equation}
a(t)=a_{bg} \left[1-\frac{\Delta}{B(t)-B_\infty}\right]. \label{eq:a}
\end{equation}
This change of the scattering length in time has an average $\bar{a}$, a minimum $a_-$, a maximum $a_+$, and is only approximately sinusoidal.

For all numerical simulations, we thus use a time-dependent interaction strength [cf. Eq.~(2) of the main text]:
\begin{eqnarray}
W(z_i-z_j,t) &=& g(t) \delta(|z_i - z_j|), \nonumber\\
&=& g_0 \left[\beta_1  (-1 + (\beta_2 - \beta_3 \sin(\Omega t)))^{-1}\right]  \delta(|z_i - z_j|).
\label{TDint}
\end{eqnarray}
We collect the parameters $\beta_i$ and all other parameters in the equations of this subsection in Table~\ref{tab:pars}. 
\begin{table}[!b]
	\begin{tabular}{| c | c |}\hline
		Parameter & value \\ \hline
		$\beta_1$ & $\vert \frac{a_{bg}}{\bar{a}} \vert = - \frac{\beta_2}{(\beta_2-1)}$ \\ \hline
		$\beta_2$ & $\vert \frac{\bar{B} - B_\infty}{\Delta} \vert $ \\ \hline
		$\beta_3$ &  $\vert \frac{\Delta B}{\Delta}\vert$  \\ \hline
		$a_{bg}$ & $-24.5 a_0$ \\ \hline
		$\bar{a}$ & $7.9 a_0$ \\ \hline
		$a_+$ & $20 a_0$ \\ \hline
		$a_-$ & $0.7 a_0$ \\ \hline
		$\Delta$ & $-192.3$G \\ \hline
		$\Delta B$ & $-41.3$G \\ \hline
		$\bar{B}$ & $590.9$G \\ \hline
		$B_\infty $ & $736.8$G \\ \hline
		$g_0(N-1)$ & $357$ \\ \hline
	\end{tabular}
	\caption{Parameters in our interaction model. Here, $a_0$ is the Bohr magneton and $g_0$ is the dimensionless interaction strength for a system of $N_{exp}=5.7\times 10^5$ lithium-$7$ atoms in a close-to-one-dimensional trap with radial frequency $\omega_r/(2\pi)=254$Hz and longitudinal frequency $\omega_z/(2\pi)= 8$Hz.
	}
	\label{tab:pars}
\end{table}

\subsection{Numerical parameters} 
We use the MCTDH-X software \cite{ultracold} to solve the MCTDHB equations of motion for the 1D problem of $N=10000$ trapped particles with the Hamiltonian of Eqs.~(1)-(2) of the main text. 
We first find the ground state $\Psi_0$ with no time-modulation and then propagate it in real time for $\Omega \in [0,10]$. We solve the MultiConfiguration Time-Dependent Hartree for Bosons (MCTDHB) equations for $M=2$ orbitals. 
For the parameters chosen the time unit scales to $\tau=2ms$ and the length unit is $L=4.3\mu{}m$. The problem is numerically highly demanding and the error tolerance requested is extremely high ($10^{-11}-10^{-10}$). Several different numerical integrators have been used (Runge-Kutta, Adams-Bashforth-Moulton, Bulirsch-Stoer) of variable time step, and the results are found to diverge within an accepted error of a few $\%$. Asymmetries seen in the density of the second orbital are due to the high stiffness of the coupled partial differential MCTDHB equations of motion, induced by the rapid oscillation of the two-body local potentials.

\section{Perturbative expansion \label{Sec:Dyson1}}
To understand the dynamical behavior -- adiabatic evolution and granulation -- observed in experiment~\cite{GRAINS} and the main text, we now discuss analytically the Gross-Pitaevskii equation.
To make the analytical considerations feasible, we approximate the interaction strength in Eq.~\eqref{TDint} by a sinusoidal modulation, $g(t)\approx \tilde{g}(t)$, throughout this Section:
\begin{equation}
\tilde{g}(t) = g_0(1+A\sin(\Omega t)) \label{eq:modint}
\end{equation}
We start and underpin how the term ``adiabatic evolution'' is apt in subsection [\ref{sec:adiabatic}]. We develop a Dyson series \cite{Joachain1975} for the initial dynamics of the system with a modulated interparticle interaction [\ref{sec:Dyson2}] that shows how time-dependent interparticle interactions first generate modulations in the phase which are proliferated to the density in time. As a result, the density thus forms so-called \emph{grains}.

\subsection{Adiabatic evolution within the Thomas-Fermi approximation}\label{sec:adiabatic}
An adiabatic time-evolution for the density can be obtained by considering a Thomas-Fermi (TF) profile, found to be valid for the experimental parameter range studied in the main text and Ref.~\cite{GRAINS}. By inserting the modulated interparticle interaction defined in Eq.~\eqref{eq:modint} above into the TF density, we obtain a time-dependent TF profile $\rho_{TF}(\vec{r};t)$:
\begin{equation}
 \rho_{TF}(\vec{r};t) = \frac{\mu - V(\vec{r})}{\tilde{g}(t)} = \frac{\mu-\frac{1}{2}\omega^2 r^2}{g_0 (1+ A\sin(\Omega~t))} \simeq \rho_{TF}(\vec{r};t_0) \left( 1-A\sin (\Omega t) \right); \quad \quad t \ll 1 .
\end{equation}
Essentially, if the Bose-Einstein condensate follows the time-dependence of the interaction strength $\tilde{g}(t)$ adiabatically, then its density $\rho$ keeps its initial functional TF profile and oscillates in phase with $\tilde{g}(t)$.

\subsection{The emergence of granulation \label{sec:Dyson2}}
To describe the onset of granulation one needs to go beyond the adiabatic approximation in the previous subsection. 
Here we do so, and apply time-dependent perturbation theory to the 1D ($\vec{r}\equiv z$) GP equation:
\begin{equation}
i \partial_t \phi(z,t) =   \left( -\frac{1}{2}\partial^2_z + V(z) + g_0\vert \phi \vert^2 +g_0 A\sin(\Omega t)\vert \phi \vert^2   \right) \phi(z,t). \label{eq:TDGP}
\end{equation}
For simplicity (in analogy to Ref.~\cite{alon:07}), we absorbed the chemical potential $\mu$ as a phase term into the state $\phi(z,t)$.
In Eq.~\eqref{eq:TDGP}, we have split the time-dependent and time-independent parts of the interaction $\tilde{g}(t)$, because we plan to treat the time-dependent term $\propto g_0 A$ as the perturbation in the following. We thus split the Hamiltonian in Eq.~\eqref{eq:TDGP} as follows
\begin{eqnarray}
\hat H &=&  \hat H^{(0)} + \hat H^{(1)} , \nonumber\\
      \hat{H}^{(0)}  &=& -\frac{1}{2}\partial^2_z + V(z) + g_0\vert \phi(z,t) \vert^2, \nonumber \\
            \hat{H}^{(1)}  &=& +g_0 A\sin(\Omega t)\vert \phi (z,t) \vert^2. \label{eq:HGP}
\end{eqnarray}
We are going to consider the state $\phi(z,0)$ to be an eigenfunction of $H^{(0)}$ [and thereby also of $\hat{H}(t=0)$] with eigenvalue $E_0$ and consider $H^{(1)}$ as a time-dependent perturbation to $\phi(z,0)$. 

In the \emph{interaction picture}, the propagator $U(t,t_0)$ that evolves the initial state $\phi(z,t_0)$ to $\phi(z,t>t_0)$ is formally given by
\begin{equation}
 \phi (z,t) = U(t,t_0) \phi (z,t_0),\label{eq:evolution}
\end{equation}
and
\begin{equation}
 U(t,t_0) = \mathcal{T} \exp \left( -i \int_{t_0}^t \diff\tau H^{(1)}(\tau) \right).
\end{equation}
Here $\hat{\tau}$ is the time-ordering operator. In the following, we focus on a time step $t$ that is sufficiently small to ignore the time-dependence of the term $\vert \Phi (z,t)\vert^2$ in $H^{(0)}$ and use $H^{(0)}\approx -\frac{1}{2}\partial^2_z + V(z) + g_0\vert \phi(z,t_0) \vert^2$ [cf. Eq.~\eqref{eq:HGP}].
It then holds that the state at time $t$ is:
\begin{equation}
\phi(z,t) = \sum_{n=0}^\infty\frac{(-i)^n}{n!} \left( \prod_{k=1}^n \int_{t_0}^{t}\diff{}t_k \right) \hat\tau
                  \left\lbrace \prod_{k=1}^n e^{i H^{(0)} t_k} H^{(1)} e^{-i H^{(0)} t_k} \right\rbrace  \phi(z,t_0).
\label{EqDysonPhi}
\end{equation}
To investigate the emergence of granulation, we consider small a step of time $t_0\equiv0\rightarrow t_1$ that is sufficiently small for us to consider only the first term of Eq.~\eqref{EqDysonPhi}. The integral on $H^{(1)}$ in Eq.~\eqref{EqDysonPhi} can easily be evaluated and we obtain:
\begin{eqnarray}
 \phi (z,t_1)  &\approx& \exp \left[- i \frac{g_0 A}{\Omega} \vert \phi(z,0) \vert^2 (\cos(\Omega t_1) - \cos(\Omega t_0) ) -i E_0 (t_1 - t_0)\right] \phi (z,0) \nonumber \\
 &\stackrel{t_0\equiv 0}{=}& \exp \left[- i \frac{g_0 A}{\Omega} \vert \phi(z,0) \vert^2 ( \cos(\Omega t_1) -1 ) -i E_0 t_1\right] \phi (z,0) \nonumber \\ 
 &=& \exp (-i \td(z,0,t_1) )  \phi(z,0).
\label{Order1}
\end{eqnarray}
Here, in the last line the shorthand $\td(z,t_0,t_1)$ was introduced:
\begin{eqnarray}
 \td(z,t_0,t_1) &=& E (t_1 - t_0) + \frac{g_0 A}{\Omega} \vert  \phi(z,t_0) \vert^2  [ \cos ( \Omega t_1 ) - \cos(\Omega t_0)  ]\nonumber \\
 &\stackrel{t_0\equiv 0}{=}&  E_0 t_1 + \frac{g_0 A}{\Omega} \vert  \phi(z,0) \vert^2  [ \cos ( \Omega t_1 ) - 1 ] \label{eq:theta}
\end{eqnarray}
The equations Eq.~\eqref{Order1} and Eq.~\eqref{eq:theta} suggest that the initial consequence of the modulation of the interaction strength is a modulation of the phase of the state. Moreover, $\td(z,t)$ represents a first-order approximation to the time-evolution operator [cf. Eq.~\eqref{eq:evolution}]. 
We will use this finding in what follows; in our approximation, the interaction term in the GPE with the time-independent interaction strength, is unaffected by the initial phase evolution:
\begin{equation}
g_0 \vert \phi(z,0) \vert^2= g_0 \vert \phi(z,t_1) \vert^2.
\end{equation} 
Moreover, the functional shape of the time-evolution operator of $H^{(1)}$ stays the same. 

In order to show that the initial phase modulation is transformed to density undulation as time proceeds, we now analyze the state in Eq.~\eqref{eq:theta} using the hydrodynamic representation of the GPE~\cite{PethickSmith}:
\begin{eqnarray}
\frac{\partial(\rho)}{\partial t} &=& -\nabla\cdot(\rho\nabla S), \label{eq:RHO_EOM} \\
\frac{\partial S}{\partial t}    &=& -\frac{1}{2 \sqrt\rho}  \nabla^2\sqrt\rho + \frac{1}{2}|\nabla S|^2 + V(\vec{r}) + g \rho  \label{GP2}
\end{eqnarray}
We continue by considering Eq.~\eqref{eq:RHO_EOM} for a small step $t_1\rightarrow t_2$. Using a leading-order Taylor expansion, we write 
\begin{equation}
\rho(z,t_2) \approx \rho(z,t_1) + (t_2-t_1)\left.\frac{\partial\rho(z,t)}{\partial t}\right|_{t=t_1}
\end{equation}
Now, we insert the right-hand side of Eq.~\eqref{eq:RHO_EOM} and use our result that the modulated interactions result in a phase modulation $\theta(z,0,t_1)$ for sufficiently small $t_1$ [implying $\rho(z,t_1)=\rho(z,0)=\vert \phi(z,0) \vert^2$ and $S(z,t)=\theta(z,t_0,t)$]:
\begin{equation}
\rho(z,t_2) \approx \rho(z,0) + (t_2-t_1) \left. \left[ \partial_z \left(\rho(z,t) \partial_z \td(z,t_0,t) \right) \right] \right|_{t=t_1} \label{eq:RHOt2}
\end{equation}
Formally, we obtain for the spatial derivative of $\td(z,t_0=0,t)$:
\begin{equation}
\frac{\partial}{\partial z} \td (z,0,t) =\frac{2 g_0 A}{\Omega} \vert  \phi(z,0) \vert \vert  \phi(z,0) \vert' [ \cos ( \Omega t ) - 1  ].  \label{eq:dztheta}
\end{equation}
Here, and in the following we use the $'$ to symbolize the spatial derivative. For the derivative of $\rho(z,t)=\vert \phi(z,t) \vert^2$ with respect to $z$, we find:
\begin{equation}
\frac{\partial}{\partial z} \rho(z,t) = 2 \vert  \phi(z,t) \vert \vert  \phi(z,t) \vert'. \label{eq:dzRHO}
\end{equation}
We move on and insert the derivatives in Eq.~\eqref{eq:dztheta} and \eqref{eq:dzRHO} into Eq.~\eqref{eq:RHOt2}:
\begin{eqnarray}
\rho(z,t_2) &\approx& \rho(z,0) + (t_2-t_1) \frac{\partial}{\partial z} \left[\rho(z,t) \frac{2 g_0 A}{\Omega} \vert  \phi(z,0) \vert \vert  \phi(z,0) \vert' [ \cos ( \Omega t ) - 1  ]\right] \Bigg|_{t=t_1}\\ \nonumber
&=& \rho(z,0) + (t_2-t_1)  [ \cos ( \Omega t ) - 1 ] \left[ \frac{4 g_0 A}{\Omega} \vert  \phi(z,t) \vert \vert  \phi(z,t) \vert' \vert  \phi(z,0) \vert \vert  \phi(z,0) \vert' \right. \\ \nonumber
& & \qquad \qquad  \qquad \qquad  \qquad \qquad+ \left. \rho(z,t) \frac{2 g_0 A}{\Omega} \left( \left[\vert \phi(z,0) \vert'\right]^2 + \vert  \phi(z,0) \vert \vert  \phi(z,0) \vert'' \right)  \right] \Bigg|_{t=t_1} \\
&=& \rho(z,0) \left[ 1+ (t_2-t_1) \frac{6 g_0 A}{\Omega}  [ \cos ( \Omega t_1 ) - 1 ]\left[ \vert \phi(z,0) \vert'  \right]^2 \right] \equiv \rho(z,0)\left[1+ T(z,\Delta t) \right]. \label{eq:RHOmod}
\end{eqnarray}
In the last line, we used that $\Delta t= t_2 - t_1$, $\rho(z,t_1)=\rho(z,0)$, inserted $t=t_1$, and neglected $ \vert  \phi(z,0) \vert''$ due to the initial Thomas-Fermi density profile. Moreover, we defined the density modulation $T(z,\Delta t)$. Importantly, Eq.~\eqref{eq:RHOmod}, to leading order, shows explicitly that a phase modulation results in a density modulation as time is proceeding. 

We note, that our approximations here represent a minimal and qualitative example. 
If, for instance, terms beyond the first order in the time-evolution operator, Eq.~\eqref{EqDysonPhi}, are considered, this results in the emergence of products of multiple $\cos{\Omega t}$ terms in the phase modulation [cf. Eq.~\eqref{eq:theta}]. Such a more complicated phase modulation would result in a much richer density modulation with multiple frequencies. A similar reasoning applies to our leading-order treatment to obtain Eq.~\eqref{eq:RHOmod}.
Evidently, the present perturbative treatment can thus  only be applied strictly to a small initial period of time. 

Additionally, due to the emergent correlations in the state, the phase and density evolutions of multiple orbitals needed to be considered instead of our above single-orbital/Gross-Pitaevskii analysis. 
Such a many-body perturbative treatment could be achieved using the MCTDHB orbital equations of motion~\cite{alon:07}. In the case of $M$ orbitals $\phi_1(z,t),...,\phi_M(z,t)$, one obtains density modulations $\mathcal F_{k}(t_2,g_0,\Omega,\phi_1(z,0),...,\phi_M(z,0))$,$k=1,...,M$, respectively. Importantly, in the many-body case, the phase modulations in leading order and the resulting time-proliferated density modulations order depend on all of the considered orbitals $\phi_1(z,0),...,\phi_M(z,0)$.
The complexity of the phase and density evolution is thus substantially increased, if more than a single orbital is considered.

\section{One- and many-body phase and their effect on correlations}\label{Sec:Phase}
In this section, we point out the fundamental relations of phases for the mean-field and the many-body cases in conjunction with the phases of the respective one-body basis or orbitals. 
As we show below, the phase of a many-body state becomes \textit{exponentially} more complicated, as more particles do not occupy a single orbital, a new measure for the phase for the many-body case is needed. Following Refs.~\cite{Stimming2011}, a good estimate of the relative phase which is measured in time-of-flight sequences can be obtained from the first order correlation function. 
For simplicity, we omit specifying the time-dependence of the quantities in this subsection.
The phase of a many-body state $\Phi(\vec{r}_1,...,\vec{r}_N)$ can be expressed by decomposition of the state into modulus and phase as follows:
\begin{equation}
 \Phi(\vec{r}_1,...,\vec{r}_N) = \vert \mu \vert e^{i S (\vec{r}_1,...,\vec{r}_N)}. 
\end{equation}
Let us now first consider the phase in the mean-field case:
\begin{eqnarray}
 \Phi_{MF}(\vec{r}_1,...,\vec{r}_N)&=& \mathcal{N} \prod_{k=1}^N \phi_{GP} (\vec{r}_k) = \vert \mu_{MF}(\vec{r}_1,...,\vec{r}_N) \vert e^{iS_{MF}(\vec{r}_1,...,\vec{r}_N)} \\
 \vert \mu_{MF} \vert &=& \prod_{k=1}^N \vert \mu_{GP}(\vec{r}_k) \vert\; \text{and} \; S_{MF} = \sum_{k=1}^N S_{GP}(\vec{r}_k), \; \text{where} \\
 \phi_{GP}(\vec{r}) &=& \vert \mu_{MF}(\vec{r}) \vert e^{i S_{GP} (\vec{r}) }. 
\end{eqnarray}
In the present mean-field case, therefore, the phase of the many-body wavefunction $S_{MF}(\vec{r}_1,...,\vec{r}_N)$ is obtained by summing the phases of the GP orbital, $S_{GP}(\vec{r}_j)$ for $j=1,...,N$. 

For a many-body state built from two single-particle wavefunctions or orbitals $\phi_1(\vec{r}),\phi_2(\vec{r})$, on the other hand, one obtains the phase
\begin{eqnarray}
 \Phi_{MB}(\vec{r}_1,...,\vec{r}_N) &=& \sum_{k=0}^N C_k  \vert N-k , k \rangle, \label{PSIMB} \\
  &=&  \vert \mu_{MB}(\vec{r}_1,...,\vec{r}_N) \vert e^{iS_{MB}(\vec{r}_1,...,\vec{r}_N)}.
\end{eqnarray}
In the following, we shall show the dependence of the many-body phase $S_{MB}$ on the single particle phases defined by
\begin{equation}
 \phi_1 (\vec{r}) = \vert \mu_1 \vert e^{i S_1(\vec{r})}\; \text{and} \; \phi_1 (\vec{r}) = \vert \mu_2 \vert e^{i S_2(\vec{r})}.
\end{equation}
To proceed, we first express the two-orbital many-body basis state $\vert N-k,k \rangle$ in first quantized notation scrutinizing the symmetrization operator $\hat{\mathcal{S}}$:
\begin{equation}
 \vert N-k , k \rangle= \hat{\mathcal{S}} \Bigg[\prod_{i=1}^{N-k} \phi_1(\vec{r}_{i}) \prod_{j=N-k+1}^N \phi_2(\vec{r}_j) \Bigg]. \label{conf1q}
\end{equation}
It is now straightforward to express the configuration $\vert N-k,k \rangle$ in terms of the moduli and phases of the single-particle states $\phi_1(\vec{r}),\phi_2(\vec{r})$:
\begin{eqnarray}
 \vert N-k , k \rangle &=&  \hat{\mathcal{S}} \Bigg[ \prod_{i=1}^{N-k}  \vert \mu_1(\vec{r}_i)  \vert e^{i S_1(\vec{r}_i)}  \prod_{j=N-k+1}^N \vert \mu_2(\vec{r}_j) \vert e^{i S_2(\vec{r}_j)} \Bigg], \\
 &=& \hat{\mathcal{S}} \Bigg[  \exp\left(i \lbrace\sum_{i=1}^{N-k} S_1(\vec{r}_i) + \sum_{j=N-k+1}^N S_2(\vec{r}_j)\rbrace \right) \prod_{i=1}^{N-k} \vert \mu_1(\vec{r}_i) \vert  \prod_{j=N-k+1}^N \vert \mu_2(\vec{r}_j) \vert \Bigg].
\end{eqnarray}
With this result, we can now write down the many-body phase $S_{MB}$:
\begin{equation}
S_{MB}(\vec{r}_1,...,\vec{r}_N) = \sum_{k=0}^N S_{C_k} \hat{\mathcal{S}} \Bigg[  \sum_{i=1}^{N-k} S_1(\vec{r}_i) + \sum_{j=N-k+1}^N S_2(\vec{r}_j)\Bigg]. 
\label{MBS}
\end{equation}
Here, $S_{C_k}$ is a scalar phase factor obtained from the decomposition of the coefficient $C_k$ onto his modulus and phase, $C_k= \vert C_k \vert \exp( i S_{C_k} )$.
Let us remind here, that the symmetrization operator $\hat{\mathcal{S}}$ applied to a product of single-particle states $\phi_1(\vec{r}),...,\phi_M(\vec{r})$ creates a sum containing $n_1! n_2! \cdots n_M!$ terms. This means that the term $\hat{\mathcal{S}}\left[ \cdot \right]$ in Eq.~\eqref{MBS} contains $(n-k)!n!$ sums of the $N$ one-body phases $S_1,S_2$. Hence, for every significantly contributing coefficient in a many-body state [cf. Eq.~\ref{PSIMB}] one therefore gets an exponentially large sum of one-body phases $S_1,S_2$. Moreover, the number of contributing terms to the many-body phase $S_{MB}$ is exponentially larger than the number of terms contributing to the phase in the mean-field case $S_{MF}$ even in the case that only a small number of coefficients $C_k$ is contributing to the wavefunction $\Phi_{MB}$. 
We infer that the many-body phase $S_{MB}(\vec{r}_1,...,\vec{r}_N)$ will become chaotic very quickly when more than one configuration is contributing in a case that either $S_1$, or $S_2$, or both, do have a nontrivial structure. Due to the relation pointed out in Ref.~\cite{Stimming2011}, one can quantify the \textit{relative phase} of two points $\vec{r}$ and $\vec{r}'$ in the system by the first order correlation function 
\begin{equation}
\vert g^{(1)}(\vec{r}) \vert = \Bigg| \frac{\rho^{(1)}(\vec{r},-\vec{r})}{\sqrt{\rho^{(1)}(\vec{r},\vec{r})\rho^{(1)}(-\vec{r},-\vec{r})}} \Bigg| = \Bigg| \frac{\langle \Psi \vert \hat{\Psi}^\dagger(\vec{r}) \hat{\Psi}(-\vec{r}) \vert \Psi \rangle}{\sqrt{\langle \Psi \vert \hat{\Psi}^\dagger(\vec{r}) \hat{\Psi}(\vec{r}) \vert \Psi \rangle \langle \Psi \vert \hat{\Psi}^\dagger(-\vec{r}) \hat{\Psi}(-\vec{r}) \vert \Psi \rangle}}  \Bigg|.
\end{equation}

\section{Entropy in the many-body dynamics} \label{Sec:Entropy}
\begin{figure}
\includegraphics[width=\linewidth]{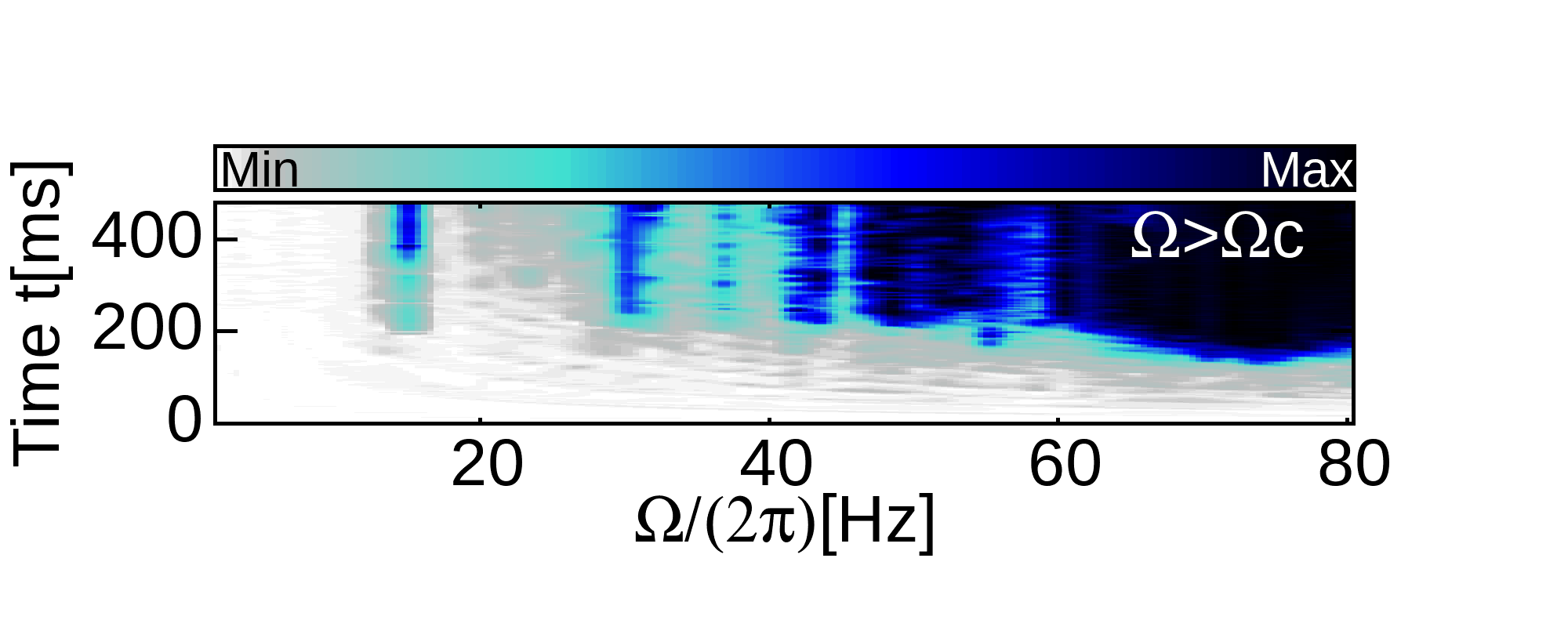}\\
\includegraphics[width=\linewidth]{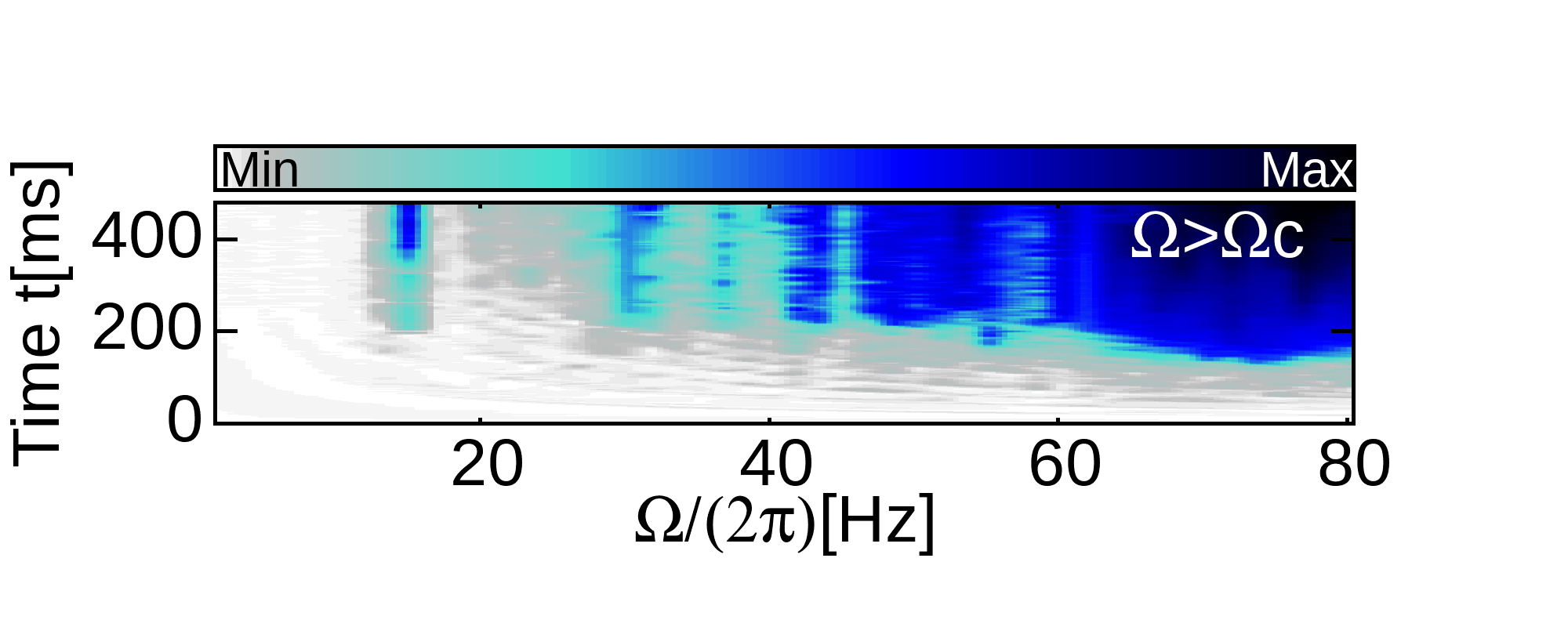}
\caption{Time evolution of the one-body (top) and two-body (bottom) entropy [Eq.~\eqref{eq:S12}] as a function of modulation frequency $\Omega$. Granulation, fragmentation (Fig. 1 in the main text) and entropy follow the same pattern and emerge side-by-side.}
\label{Fig:Entropy}
\end{figure}

The evolution of fragmentation and granulation is seen to be in sync from Fig.~1 of the main text. Here we show, that the entropy of the one-body and two-body density matrix (cf. Refs.~\cite{chakrabarti:15,roy:18}), 
\begin{equation}
\mathbf{S}_{\rho^{(1,2)}} = - \sum_k \alpha^{(1,2)}_k ln \frac{\alpha^{(1,2)}_k}{N}, \label{eq:S12}
\end{equation}
exhibits a similar behavior. Here, $\alpha^{(1)}_k$ are the eigenvalues of the one-body density,
\begin{equation}
\rho^{(1)}(z,z')=\sum_k \alpha^{(1)} \phi_k^{*,(NO)}(z',t)\phi_k^{(NO)}(z,t) \label{eq:NOs}
\end{equation}
and $\phi_k^{(NO)}$ denotes its eigenfunctions. The $\alpha^{(2)}_k$ are the eigenvalues of the two-body density and are defined analogously for $\rho^{(2)}$, see Ref.~\cite{sakmann:08}.
The time evolution of $\mathbf{S}_{\rho^{(1,2)}}$ is plotted as a function of modulation frequency $\Omega$ in Fig.~\ref{Fig:Entropy}.

\section{Higher-order correlation functions}\label{Sec:HC}
\begin{figure}
	\begin{center}
\includegraphics[width=0.5\linewidth]{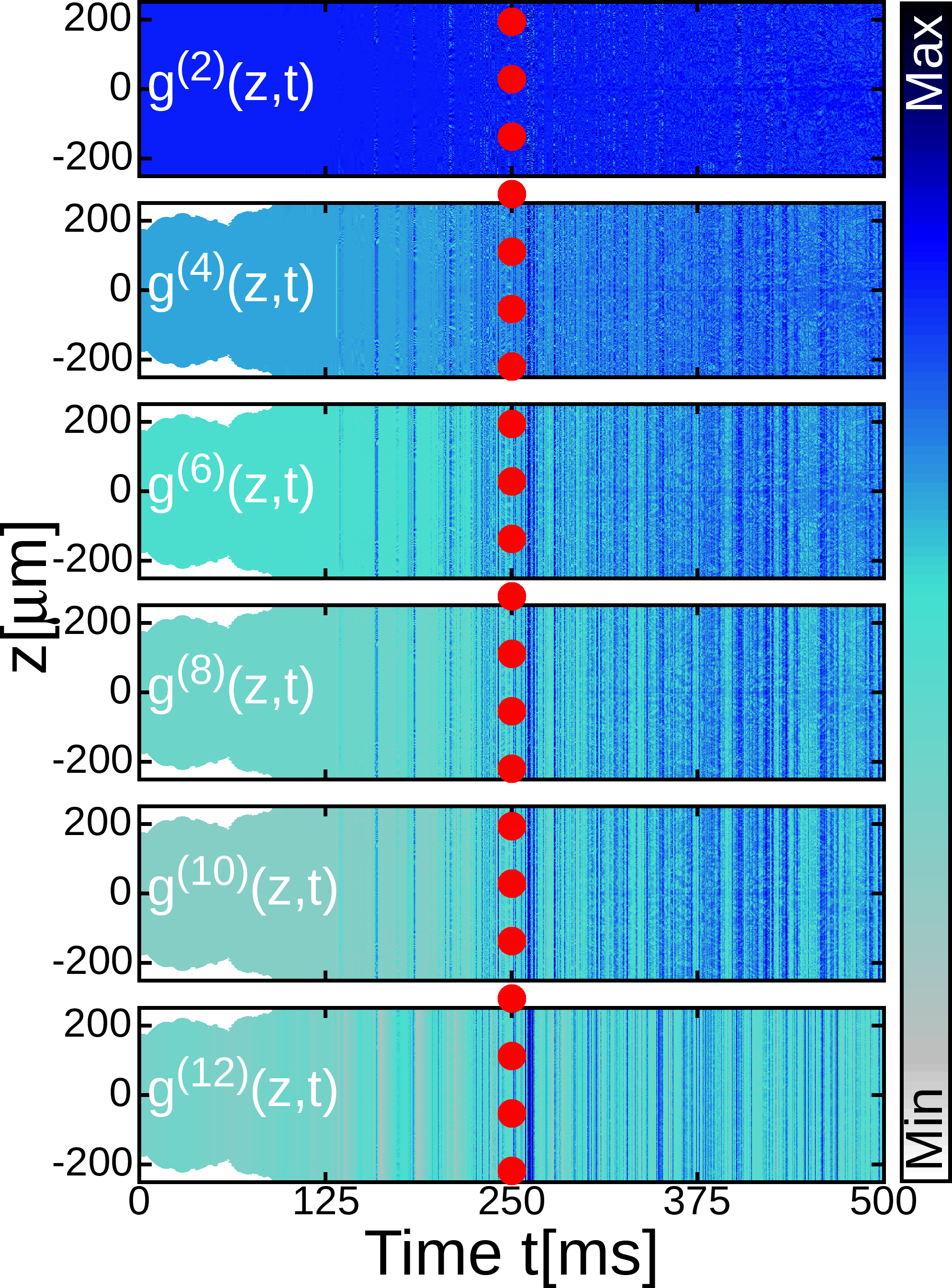}
	\end{center}
\caption{Higher-order correlation functions as a function of time for a granulated Bose-Einstein condensate. As a consequence of the many-body phase evolution (Sec.~\ref{Sec:Phase}), higher-order correlations emerge as quantified here via the correlation function with the trap center, $g^{(p)}$, for $p\leq12$ at $\Omega=10\omega_z$ [Eq.~\eqref{Eq:HC}]. See text for discussion.}
\label{Fig:HC}
\end{figure}

To complement our argument that granulation in BEC is a result of exponentially many phase terms affecting the time-evolution, we show here that the emergence of higher-order correlations is triggered by its emergence.
We focus on correlations of the atoms with the centre of the trap at $z=0$:
\begin{equation}
  g^{(p)}(z) =\frac{\rho^{(p)}(z,0,...,0,z'=z,0,...,0)}{\rho^{(1)}(z,z)\left[\rho^{(1)}(0,0)\right]^{(p-1)}}.\label{Eq:HC}
\end{equation}
This correlation function describes correlations when $p-1$ bosons are fixed at $z=0$, see Fig.~\ref{Fig:HC} for a plot.
Our findings underpin the analytical and numerical evidence that granulation is a state that is made persistent via quantum correlations.
At higher orders $p$, we note that an irregularly striped pattern emerges in the evolution of $g^{(p)}$. The reason is our choice to pin $p-1$ particles at the center of the trap -- for increasing orders $p$, we probe correlations of more and more ``remote'' areas in the many-particle Hilbert space. At $p=12$ (lower-most plot in Fig.~\ref{Fig:HC}), for instance, we plot the likelihood that a particle detected at position $z$ is phase coherent \emph{given that} $11$ particles were found at $z=0$. If, however, the probability to find $11$ or $p-1$ particles in the trap center is very little, because (due to the granularity of the state) $\rho^{(p-1)}(0,...,0)$ is small, then our measure $g^{(p)}$ will show the absence of correlations with a turquoise ($g^{(p)}\approx 1$) stripe. When, in turn, a grain with more than $p-1$ particles is present at $z=0$, strong correlations across the entire many-body state are signaled by a blue/black stripe ($g^{(p)}\gg 1$).